\documentclass[11pt]{article}

\title{On the Gompertzian dynamics of growth and self-organization}
\author{Marcin Molski and Jerzy Konarski\\
\phantom{x}\\
Department of Theoretical Chemistry, Faculty of Chemistry\\
A. Mickiewicz University of Pozna\'n,\\
ul. Grunwladzka 6, PL60-780 Pozna\'n, Poland}
\date{}
\begin{document}

\maketitle

\begin{abstract}

Comment on the Waliszewski's article "A principle of fractal-sto-cha\-stic dualism and Gompertzian 
dynamics of growth and self-orga\-ni\-zation" (BioSystems 82 (2005)61-73) is presented. 
It has been proved that the main idea of this work that Gompertzian dynamics is governed by the 
Schr\"{o}dinger-like equation including anharmonic Morse potential has been already introduced by 
Molski and Konarski in 2003. Some inconsistencies and mathematical errors in the Waliszewski's 
model are also pointed out. 

\end{abstract}

In a recent article published in BioSystems \cite{Wali} an idea has been developed 
that the Gompertzian dynamics of growth \cite{Gomp}  is governed by the Schr\"o\-dinger-like equation 
including Morse potential widely applied in spectroscopy of diatomic systems \cite{Molski, Molski1}. 
In particular the equation of growth has been proposed \cite{Wali}
\begin{equation}
-\frac{1}{b^2}\frac{d^2f(t)}{dt^2}+\left[ae^{-bt}-\frac12\right]^2f(t)=\frac{1}{4}f(t),
\label{A1}
\end{equation}
in which 
\begin{equation}
f(t)=e^{{a}\left(1-e^{-bt}\right)}
\label{A2}
\end{equation}
is one of the representations of the Gompertz function. Although the author published Eq.(\ref{A1}) 
as his finding, an identical equation 
\begin{equation}
-\frac{d^2G(t)}{dt^2}+\frac{a^2}{4}\left(1-\frac{2b}{a}e^{-
at}\right)^2G(t)=\frac{a^2}{4}G(t)
\label{eq5}
\end{equation}
has been already introduced by Molski and Konarski in 2003 \cite{Molski2} for Gompertz function specified in the form
\begin{equation}
G(t)=G_0e^{\frac{b}{a}\left(1-e^{-at}\right)},
\label{A4}
\end{equation}
in which $a$ and $b$ are retardation and regression rate constants. 
To compare Eqs.(\ref{eq5}) and (\ref{A1}) we rewrite the former to the form
\begin{equation}
-\frac{1}{a^2}\frac{d^2G(t)}{dt^2}+\frac{1}{4}\left(\frac{b}{a}e^{-at}-\frac12\right)^2G(t)=\frac{1}{4}G(t).
\label{A3}
\end{equation}
A look into Eq.(\ref{A1}) and Eq.(\ref{A3}) reveals that these equations are identical to within constants  
$a$ and $b$ redefined in the Gompertz function (\ref{A2}) by the author \cite{Wali}.

The author \cite{Wali} (p.65) has posed an important question whether the relationship between the operator 
differential equation (\ref{A1}), the Gompertz function (\ref{A2}) and the Morse-like anharmonic potential represents
a universal relationship underlying the growth and self-organization of dynamics of a supramolecular cellular system or 
it is a kind of mathematical illusion. To unswer this question a space-dependent Morse \cite{Morse} function has been taken
into  consideration \cite{Wali}
\begin{equation}
U(x(t))=D\left(1-e^{-bx(t)}\right)^2
\label{N1}
\end{equation}
in which $D$ and $b$ are dissociation and range potential constants whereas distance $x(t)$ is expressed as a 
continuous function of time. According to the author \cite{Wali} (Eq.(11)) the potential (\ref{N1}) can be related to the Gompertz
function (\ref{A2})
\begin{equation}
f(t)=e^{{a}\left(1-e^{-bt}\right)}=e^{\sqrt{U(t)}}
\label{N2}
\end{equation}
at the assumptions 
\begin{equation}
x(t)=t\hskip1cm a=\sqrt{D}.
\label{N3}
\end{equation}
The assumption $a=\sqrt{D}$ leading to Eq.(\ref{N2}) has no physical meaning as the constant $a$ in the 
Gompertz function (\ref{A2}) is dimensionless whereas $D$ in Eq.(\ref{N1}) stands for dissociation energy of the system
\cite{Morse}. Hence, proposed by the author relation (\ref{N2}) linking
Gompertz function (\ref{A2}) and the Morse potential  (\ref{N1}) is a mathematical illusion.

Another inconsistency appears in the derivation of the anharmonic potential for the expanding metabolizing cellular mass
\cite{Wali} (Eq.(20) ) from the Maclaurin expansion of the Morse function (\ref{N1})
\begin{equation}
U(x(t))= U(x(t)=0) +\left(\frac{\partial U}{\partial x(t)}\right)_{x(t)=0}x(t)+\frac12\left(\frac{\partial^2U}{\partial x(t)^2}\right)_{x(t)=0}x(t)^2+......
\label{N61}
\end{equation}
which is reduced to the second order harmonic term
\begin{equation}
U(x(t))= \frac12\left(\frac{\partial^2U}{\partial x(t)^2}\right)_{x(t)=0}x(t)^2.
\label{N6}
\end{equation}
According to the author, $x(t)$ has been considered in the form \cite{Wali}
\begin{equation}
x(t)=e^{\frac{a}{b_s}(1-e^{-bt})+\frac{1}{b_s}\ln{\frac{V_0}{a_s}}},
\label{N21}
\end{equation}
Here $a_s$ and $b_s$ stand for scaling coefficient and spatial 
fractal dimension, respectively, whereas $V_0$ is the initial volume of cells.
After substitution Eq.(\ref{N21}) into Eq.(\ref{N6}) the latter is named by the author \cite{Wali} (p.66) {\it ...the potential energy 
of the anharmonic oscillator for interacting supramolecular cellular systems which exist and interact in fractal time-space.}
This interpretation is internally inconsistent as the second order term of the Maclaurin expansion (\ref{N61}) is
interpreted as a harmonic potential and not anharmonic one. In particular the second order derivative in Eq.(\ref{N6}) 
calculated at the fixed point $x(t)=0$ 
\begin{equation}
\left(\frac{\partial^2U}{\partial x(t)^2}\right)_{x(t)=0}=2Db^2=k
\label{N7}
\end{equation}
is the force constant of a harmonic oscillator. In view of this it is unclear: 
\begin{description}
\item[(i)] why the force constant in  \cite{Wali} (Eq.(19)) has been calculated without 
fixing $x(t)=0$, although the author uses the proper definition of $k$ given by Eq.(\ref{N7});
\item[(ii)] why in the calculation of the potential energy of the anharmonic oscillator for interacting supramolecular 
cellular systems the harmonic term Eq.(\ref{N6}) instead of the anharmonic Morse potential Eq.(\ref{N1}) has been used;
\item[(iii)] how the plot of the anharmonic potential \cite{Wali} (Eq.(20))  in dependence on the $x(t)$ variable 
can be generated \cite{Wali} (Fig.1) if $U(x(t))$ in the author's model depends both on time $t$ and $x(t)$ coordinate.
\end{description}
We conclude that the proper form of anharmonic potential intended to be derived by Waliszewski should 
be a combination of Eq.(\ref{N1}) and Eq.(\ref{N21}) yielding
\begin{equation}
U(t)=D\left(1-e^{-be^{\frac{a}{b_s}(1-e^{-bt})+\frac{1}{b_s}\ln{\frac{V_0}{a_s}}}}\right)^2.
\label{N8}
\end{equation}
Unfortunately, even the aforementioned equation is incorrect as the basic equation \cite{Wali} (Eq.(13))  
\begin{equation}
V=a_sx(t)^{b_s}=V_0e^{1-e^{-bt}}
\label{N9}
\end{equation}
employed to derive Eq.(\ref{N21}) has no physical meaning. It is as $\lim_{x(t)\to 0}V_0e^{1-e^{-bt}}=V_0$, whereas 
$\lim_{x(t)\to 0}a_sx(t)^{b_s}=0$. 

Besides the aforementioned inconsistencies one may find in \cite{Wali} numerous mathematical errors 
and other unphysical assumptions specifed below. 
\begin{description}
\item[(i)] Eq.(11) in \cite{Wali}, is not only unphysical but also contains the error: $\sqrt{D}$ should be replaced by $D$.  
\item[(ii)] The Maclaurin series given by Eq.(18) in  \cite{Wali} is incorrect as it does not contain multiplicative 
factors $1/n!$. In consequence this equation contains three mathematical errors. 
\item[(iii)]  Eq.(B.4) in  \cite{Wali} $e^{1-e^{-bt}}=a_tt^{b_t}$ has no physical meaning as $\lim_{t\to 0}e^{1-e^{-bt}}=1$, whereas   
$\lim_{t\to 0}a_tt^{b_t}=0$. In consequence the relationships (3) and (B.6) derived  in  \cite{Wali} are incorrect.
\item[(iv)] The plot in Fig.4  \cite{Wali} does not represents $F_{Gauss}(t)$ as on the plot for $t=0$,  $F_{Gauss}(0)=1$, 
whereas it should be $F_{Gauss}(0)=[1/(2.5\sqrt{2\pi})]$.  
\end{description}

\end{document}